\documentclass[amsmath,amssymb,floatfix,showpacs,twocolumn]{revtex4}
\usepackage{amsfonts}
\usepackage{graphicx}
\usepackage{dcolumn}
\usepackage{bm}
\usepackage{hyperref} 
\usepackage{latexsym}
\usepackage{float}

\def\av#1{\langle#1\rangle}

\def\annil{$A+A\to0$}
\def\coal{$A+A\to A$}

\begin{document}
\title{Diffusion-Limited One-Species Reactions in the Bethe Lattice}   

\author{Daniel ben-Avraham}
\email{benavraham@clarkson.edu}
\affiliation{Department of Physics, Clarkson University,
Potsdam NY 13699-5820, USA}
\author{M. Lawrence Glasser}
\email{laryg@clarkson.edu}
\affiliation{Department of Physics, Clarkson University,
Potsdam NY 13699-5820, USA}

\begin{abstract}
We study the kinetics of diffusion-limited coalescence, \coal, and annihilation, \annil,  in the Bethe lattice of coordination number $z$.  Correlations build up over time so that the probability to find a particle next to another varies from $\rho^2$ ($\rho$ is the particle density), initially, when the particles are uncorrelated, to $[(z-2)/z]\rho^2$, in the long-time asymptotic limit.  As a result, the particle density decays inversely proportional to time, $\rho\sim1/kt$, but at a rate $k$ that slowly decreases to an asymptotic constant value.
\end{abstract}

\maketitle

\section*{INTRODUCTION}
Diffusion-limited reactions are dominated by fluctuations at all length scales.  In lack of a coherent approach for their analysis the simplest reaction schemes, such as one-species coalescence, \coal, and annihilation, \annil, have attracted the most attention~\cite{book}. In one-dimensional space their kinetics can be analyzed exactly and is found to be {\it anomalous\/} --- at variance with the mean-field (reaction-limited) result: the density of particles decays as $\rho\sim1/\sqrt{t}$.  In two dimensions
$\rho\sim~\ln t/t$, and only in three dimensions and above does one observe the mean-field result $\rho\sim1/t$~\cite{book}.

\begin{figure}[ht]
\vspace*{0.cm}
 \includegraphics*[width=0.3\textwidth]{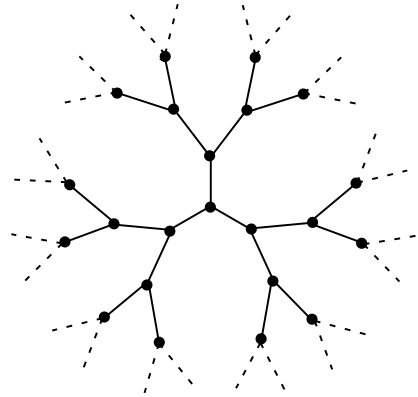}
\caption{Bethe lattice, of coordination number $z=3$.}
\label{cayley}
\end{figure}

Here we analyze coalescence and annihilation in the Bethe lattice of coordination number $z$ (Fig.~\ref{cayley}).   Our analysis, which is based on a generalization of the method of intervals~\cite{intervals,masser,lindenberg}, shows that even in these infinite-dimensional objects fluctuations play an important role,
slowing down the rate of reactions.   The correlation-free mean-field limit is achieved only as $z\to\infty$.  Other reaction schemes have been studied on the Bethe lattice~\cite{Kelbert}, including trapping~\cite{trapping}, diffusion-limited aggregation~\cite{aggregation,majumdar03} --- a process related to data compression and the Ziv-Lempel algorithm~\cite{majumdar03} --- and annihilation between immobile reactants~\cite{majumdar93}.  Diffusion in the Bethe lattice is also of interest to probabilists because of its close analogy to diffusion in hyperbolic (negative curvature) space~\cite{hyperbolic}.  From a practical viewpoint, our system models exciton annihilation in extended dendrimers.  These Cayley-tree-like macromolecules are being studied as potential nanoscale antennae, for their light harvesting capabilities~\cite{shapir,kopelman}. 

The Bethe lattice of coordination number $z$ may be obtained by starting from a single node (at shell $\ell=0$) that is connected to $z$ neighbors, at $\ell=1$.  Each of the nodes in shell $\ell>0$ is connected to $(z-1)$ nodes in shell $\ell+1$ and the process continues indefinitely (Fig.~\ref{cayley}).  When the construction stops, at shell $\ell=L$, say, the resulting graph is known as a {\it Cayley tree\/}.  The Bethe lattice is the {\it interior\/} of an {\it infinite\/} Cayley tree (i.e., it has no boundary).   
It is easy to show, for example by induction, that a connected cluster of $n$ nodes, in the Bethe lattice, has exactly $n(z-2)+2$ {\it neighbors\/} (external nodes connected to the cluster), regardless of its topology.  
The case of $z=2$ degenerates to a one-dimensional chain and has been discussed elsewhere~\cite{book,intervals}.
In the following we assume that $z\geq3$.
We further assume that each site of the infinite Cayley tree is initially occupied by a particle $A$ with probability $p$ (or empty, with probability $1-p$).  Particles hop to one of their $z$ nearest neighbors, chosen at random, at constant rate 1.  Thus, in one unit time all particles hop once, on average.
If a particle hops onto a node that is already occupied, the two particles disappear immediately, in the case of annihilation, or they merge into a single particle (at the target site), in the case of coalescence.  

\section*{SIMPLER MODELS AND APPROACHES}
\subsection*{(a) The Smoluchowski Model}
To gain intuition into the problem, let us first look at the Smoluchowski reaction model: a trap resides at the origin of the Bethe lattice, and the $A$ particles disappear as soon as they hit the trap, but otherwise do
{\it not\/} interact with one another.  This process is far simpler than coalescence or annihilation, yet it lets us peek into the nature of correlations arising because of reactions (with the trap).

Let $\rho_{\ell}(t)$ be the density of particles in shell $\ell$, at time $t$,  then,
\begin{equation}
\label{smol}
\begin{split}
&\frac{d}{dt}\rho_{\ell}(t)=\frac{1}{z}\Big(\rho_{\ell-1}-z\rho_{\ell}+(z-1)\rho_{\ell+1}\Big),\quad\ell>0,\\ 
&\rho_0(t)=0.
\end{split}
\end{equation}
The first equation reflects the fact that each site in shell $\ell>0$ is connected to just one site in shell 
$\ell-1$, and to $(z-1)$ sites in shell $\ell+1$, while all particles hop at unit rate; the second equation 
specifies the trapping reaction at the origin, $\ell=0$, as a boundary condition.  Eqs.~(\ref{smol}),
with the initial condition $\rho_{\ell}(0)=p$, admit the steady-state solution
\begin{equation}
\rho_{\ell}=p\Big[1-\Big(\frac{1}{z-1}\Big)^{\ell}\Big]\,.
\end{equation}
Thus, the density of particles next to the trap is depleted. Right next to the trap, at $\ell=1$, the density is reduced by a factor $(z-2)/(z-1)$.  The missing density decays exponentially away from the trap.  It is possible to interpret
the depleted mass, $p-\rho_{\ell}$, as a probability density, by impossing normalization: $p_{\ell}=(z-2)/(z-1)^{\ell+1}$.   From it we derive a characteristic depletion distance 
from the trap, or a reaction correlation length, $\xi=\av{\ell}=1/(z-2)$.  At $z=2$ the correlation length is infinite: 
indeed in one dimension the depletion zone near the trap expands indefinitely, as $\sqrt{t}$, and there is no steady-state.  For $z>2$ the correlation is finite, but only in the limit of $z\to\infty$ does one attain the ideal case of reaction-limited kinetics, where the correlation vanishes.   

\subsection*{(b) Mean-Field}
Consider now coalescence and annihilation in the Bethe lattice, assuming that correlations can be ignored completely.  The neglect of correlations is strictly justified only in reaction-controlled processes
and does not apply to the diffusion-controlled reactions we address in this paper, yet it yields valuable physical insights.  We shall refer to this limit as {\it mean-field\/}.  

The density of particles, $\rho(t)$, is the same in all nodes, and since particles hop at unit rate they meet one another at rate $\rho^2$, regardless of the graph coordination number $z$.  Thus, $\rho(t)$ satisfies the rate equation
\begin{equation}
\label{MFeq}
\begin{split}
&\frac{d}{dt}\rho(t)=-k\rho^2\,,\\
&\rho(0)=p\,,
\end{split}
\end{equation}
where $k=1$ for coalescence, \coal, where only {\it one\/} particle is removed upon each encounter, and $k=2$ for annihilation, \annil, where {\it two\/} particles disappear.   Eqs.~(\ref{MFeq}) have the solution
\begin{equation}
\rho(t)=\frac{1}{1/p+kt}\,,
\end{equation}
so mean-field analysis predicts an inverse-linear decay with time, $\rho\sim1/kt$, in the limit of 
$t\to\infty$.  

We see that the neglect of correlations is disastrous for the case of $z=2$ 
(one-dimensional chain), where for diffusion-limited one-species reactions $\rho\sim1/\sqrt{t}$, instead of the mean-field prediction of $1/t$.  
For $z>2$ the Bethe lattice is
infinite-dimensional: the number of nodes within $\ell$ shells increases exponentially with $\ell$ --- faster than any power (dimension) of $\ell$.  Naively, one would expect the mean-field limit to apply then, but the analysis of the Smoluchowski model, above, suggests that correlations might play a certain role, so long as $z<\infty$.
   
\section*{EMPTY INTERVALS APPROXIMATION}
We now generalize the method of empty intervals, introduced for the exact solution of coalescence on the line, to the case of coalescence in the Bethe lattice.  The generalization is not rigorous, but yields
reasonable results.
Let $E_n(t)$ denote the probability that an $n$-cluster contains no particles at time $t$.
The particle density  (i.e., the probability that a site is occupied), is
\begin{equation}
\label{rho}
\rho(t)=1-E_1(t)\;.
\end{equation}
At time $t=0$ the occupancy of nodes is independent of one another, so the probability that an $n$-cluster is empty, is
\begin{equation}
\label{IC}
E_n(0)=(1-p)^n,\qquad n=1,2,\dots
\end{equation}

Consider now changes to $E_n(t)$.  Any motion of particles {\it within\/} an $n$-cluster would not alter $E_n$, since the (non-empty) cluster cannot become empty in this way.  $E_n$ changes only when a particle hops into or out of the cluster.  To follow these changes, we require the probability $F_n$ for having an $n$-cluster that is empty, connected to a node that is occupied.  We may then write
\begin{equation}
\label{GFH}
\frac{d}{dt}E_n(t)=\frac{n(z-2)+2}{z}\Big(F_{n-1}-F_n\Big)\;.
\end{equation}
The first term on the RHS denotes the event that a particle hops out from just inside the $n$-cluster, leaving the cluster empty, thus increasing $E_n$.   The starting configuration  occurs with probability $F_{n-1}$.  The second  term, proportional to  $F_{n}$, accounts for a particle just outside of an empty $n$-cluster that hops into it: the cluster is not empty anymore and $E_n$
decreases.  Since the $n$-cluster has $n(z-2)+2$ neighbors,
and each particle hop (from one specific site to another) occurs at rate $1/z$, the overall rate for all of these
processes is $[n(z-2)+2]/z$.

\begin{figure}[ht]
\vspace*{0.cm}
 \includegraphics*[width=0.35\textwidth]{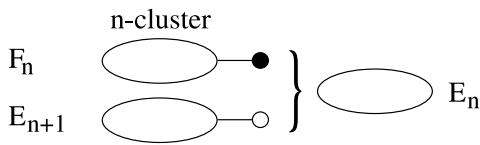}
\caption{Probability $F_n$, for finding an empty $n$-cluster (empty oval) connected to an occupied neighbor ($\bullet$).  The second symbol from the top denotes $E_{n+1}$, the probability for finding an $(n+1)$-cluster empty. The sum of the two events equals $E_n$. 
}
\label{huevos}
\end{figure}

In order to close the evolution equation~(\ref{GFH}), we need to express $F_n$ in terms of the 
$E_n$.  We note that the event that an $n$-cluster is empty (probability $E_n$) is the sum of the events that a neighboring node is also empty (probability $E_{n+1}$) and the event that the site is occupied (probability $F_n$), see Fig.~\ref{huevos}.
Hence,
\begin{equation}
F_n= E_n-E_{n+1}\,.
\end{equation}
Using this in~(\ref{GFH}), we finally obtain
\begin{equation}
\label{Gn}
\frac{d}{dt}E_n(t)=\frac{n(z-2)+2}{z}\big(E_{n-1}-2E_n+E_{n+1}\big)\;.
\end{equation}
This equation is valid for $n=2,3,\dots$  A separate analysis for the case of $n=1$ shows that it may be
subsumed in Eq.~(\ref{Gn}), with the understanding that 
\begin{equation}
\label{BC}
E_0(t)=1\;.
\end{equation}
If the system contains a finite density of particles, initially, then an infinitely large cluster can never become empty, thus
\begin{equation}
\label{BCinfty}
\lim_{n\to\infty} E_n(t)=0\,.
\end{equation}
To summarize,  we need to solve Eqs.~(\ref{Gn}), for $n=1,2,\dots$, with the boundary conditions~(\ref{BC}), (\ref{BCinfty}), and
the initial condition~(\ref{IC}).  The particle density is then obtained from~(\ref{rho}).

\begin{figure}[ht]
\vspace*{0.cm}
 \includegraphics*[width=0.35\textwidth]{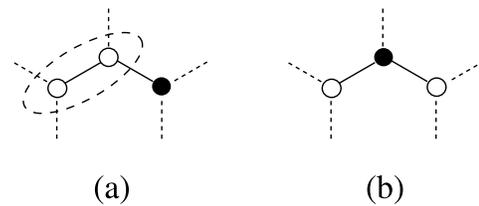}
\caption{Probability $F_2$, in the Bethe lattice of $z=3$.  The expression $F_ 2=E_2-E_3$ is correct for the configuration (a), but not for (b), where the empty sites do {\it not\/} form a connected 2-cluster (cf Fig.~\ref{huevos}).
}
\label{approx}
\end{figure}

The above argument is not exact because the relation $F_n=E_n-E_{n+1}$ is correct only when the particle is attached to the $n$-cluster through a {\it single\/} link.  The particle might be embedded
in an $(n+1)$-cluster, splitting it into two (or more) pieces.  In this case we cannot express $F_n$
in terms of the $E_n$.  In Fig.~\ref{approx} we illustrate the problem for the Bethe lattice of $z=3$, for the case of 3-clusters (the smallest cluster size where the difficulty arises).

We argue that, despite the inaccuracy, there are good reasons for pursuing the empty interval approximation:

\smallskip\noindent
$\bullet$ For $z=2$ the equations reduce to  those found for the line, where the method is {\it exact\/}.
It may be useful to think of $z$ as a continuous variable (in the spirit of analytical continuation), in which
case one expects good results for $z=2+\varepsilon$.

\smallskip\noindent
$\bullet$  The approach is exact in the limit $z\to\infty$, where correlations do not play any role.

\smallskip\noindent
$\bullet$ The approximation is good even for intermediate values of $z$.  Consider, for example, the case of $z=3$ (Fig.~\ref{approx}).  The approximation amounts to assigning the {\it same\/} weight to
configurations (a) and (b).  Since there are four ways for the particle to leave the 3-cluster in configuration (a), and only one way in configuration (b), the difference between the actual probability of (b) and the approximation ($E_2-E_3$) matters only one fifth of the time. Moreover, the analysis of the 
Smoluchowski model shows us that the correlation length is of the order of $1/(z-2)$,  that is, no more than a link's length.  Thus, the actual difference between the probabilities of (a) and (b) cannot be large.  The error involved in processing larger clusters is of even lesser consequence.

\smallskip\noindent
$\bullet$ The approximation's Eq.~(\ref{Gn}) is simple, linear and analytically tractable.  The simplest alternative
would be a Kirkwood-like truncation of the infinite hierarchy of rate equations for clusters of size $n$.
Note, however, that an $n$-cluster has $2^n$ states, so there'd be $2^n-1$ variables associated with
it, and the truncation yields a {\it non\/}-linear equation at the bottom of the hierarchy.

\subsection*{Results for Coalescence}

The empty interval equations may be tackled analytically.  For example, following a Laplace transform in the time variable the resulting system of difference equations with linear coefficients can be solved in closed form~\cite{milne}, then the transform might be inverted (see {\bf Appendix~A}).  
It is easier to focus directly on the limiting behavior.  For short times we assume the Taylor series expansion,
\[
E_n(t)=\sum_{m=0}^{\infty}a_n^{(m)}{t^m}\;.
\]
Substituting in~(\ref{Gn}) and collecting like powers of $t$ one obtains a recursion relation for the $a^{(m)}$
in terms of the $a^{(m-1)}$.  Since $a_n^{(0)}=E_n(0)$ (to satisfy initial conditions) is known, successive coefficients can be generated mechanically up to any desired order.  (This does not provide a complete solution, for convergence of the Taylor series is limited to a finite radius, $t_{\rm c}=z/(z-2)$.)  The first few terms in the particle density are
\begin{equation}
\label{rhoearly}
\rho(t)=p-p^2t+p^2\frac{1-p+pz}{z}t^2+\cdots
\end{equation}
It is interesting to compare this result to the mean-field prediction, $\rho(t)=p/(1+pt)$, that expands to
\[
\rho(t)=p-p^2t+p^3t^2+\cdots,\qquad\text{Mean-Field.}
\]
We see that the first two terms are identical, but that the quadratic term in~(\ref{rhoearly}) is larger than that in the mean-field expansion.  This reflects the fact that the initial particle distribution is 
Poissonian and devoid of correlations.   Correlations, however, do develop, resulting in the slowing down manifest in the quadratic term.

The long-time asymptotics is best analyzed by passing to a continuum limit.  As the process evolves, the density of particles drops and $n$ need be very large to find any particles within an $n$-cluster.  Under these conditions it makes sense to regard $n$ as a continuous variable ``length," $x=na$, where $a$ is the length of a link.
We then replace $E_n(t)$ with $E(x,t)$ and Eq.~(\ref{Gn}) becomes
\begin{equation}
\label{Gxt}
\frac{\partial}{\partial t}E(x,t)=\frac{z-2}{z}vx\frac{\partial^2}{\partial x^2}E(x,t)\;,
\end{equation}
where $v$ is the average speed for a particle to traverse a link.  
The boundary conditions are now $E(0,t)=1$, $\lim_{x\to\infty}E(x,t)=0$, and the particle concentration, $c(t)=\rho/a$, is, 
according to~(\ref{rho}), 
\[
c(t)=-\frac{\partial E}{\partial x}|_{x=0}\;.
\]
To reconnect with the discrete limit, we take $a=1$ and $v=1$.  

We look for a {\it scaling\/} solution, of the form 
\[
E(x,t)\to \Phi(\frac{x}{t^{\beta}})\;,\qquad\text{as\ }t\to\infty\;.
\]
Putting this back in~(\ref{Gxt}) we see that we must have $\beta=1$ and that $\Phi(x/t)$ satisfies the equation
\[
\Phi'(\xi)=-\frac{z-2}{z}\Phi''(\xi)\;,
\]
where primes denote differentiation with respect to the variable $\xi=x/t$.  The boundary conditions determine the solution
\begin{equation}
\label{Phi}
\Phi(\xi)=\exp\big(-\frac{z}{z-2}\,\xi\big)\;,
\end{equation}
and it follows that
\begin{equation}
\label{tinfty}
\rho(t)\to1/\big(\frac{z-2}{z}\,t\big)\;,\qquad\text{as\ }t\to\infty\;.
\end{equation}
Comparing the result~(\ref{tinfty}) to the 
mean-field decay rate of $\rho\sim1/t$, when $d\rho/dt=-\rho^2$, we see that the reaction rate
is effectively diminished by a factor $(z-2)/z$.  Thus, the probability to find  particles next to one another is $(z-2)/z$ times smaller than $\rho^2$.  This is reminiscent to the depletion of  the steady-state particle density near the trap in the Smoluchowski model, and is explained by the reaction mechanism: particles in proximity are disfavored in the long run, because they react more promptly and are removed from the system.  When $z\to\infty$ the correlations disappear and the mean-field limit is regained.  For $z\to2$, the case of a linear chain, the effective mean-field rate vanishes.  This is in accord with the 
known result $\rho\sim1/\sqrt{t}$~\cite{book}, which is equivalent to a radically slower reaction rate, of effective order three: $d\rho/dt=-k'\rho^3$.

\subsection*{Annihilation}

We now turn to annihilation, \annil.  Instead of $E_n$ we work with $G_n$, the probability that an $n$-cluster contains an {\it even\/} number of particles~\cite{masser,lindenberg}.  One obtains for
the $G_n$'s the same rate equations as for the $E_n$'s, Eq.~(\ref{Gn}), but with the boundary conditions; $G_0(t)=1$, $\lim_{n\to\infty}G_n(t)=1/2$.  The initial condition is too somewhat different:
The initial probability that an $n$-cluster contain an even number of particles is
\begin{equation*}
\begin{split}
&\sum_{m=0}^{\lfloor{n/2}\rfloor}\Big({n\atop2m}\Big)p^{2m}(1-p)^{n-2m}\\
&\qquad\>=\frac{1}{2}\big(p+(1-p)\big)^n+\frac{1}{2}\big(-p+(1-p)\big)^n\;,
\end{split}
\end{equation*}
or,
\begin{equation}
\label{ICg}
G_n(0)=\frac{1}{2}+\frac{1}{2}(1-2p)^n,\qquad n=1,2,\dots
\end{equation}

The short time analysis yields
\begin{equation}
\rho(t)=p-2p^2t+2p^2\frac{1+2p(z-1)}{z}t^2+\cdots,
\end{equation}
compared to the mean-field result, $\rho(t)=p/(1+2pt)$, that expands to
\[
\rho(t) = p -2p^2t +4p^3t^2+\cdots\qquad\text{Mean-Field.}
\]
The {\it early\/} buildup of correlations shows for the first time in the second-order term and depends upon the initial density $p$: the correlations slow down the rate of decline, compared to mean-field, if $p<1/2$, and accelerate the decline if $p>1/2$.  For $p=1/2$ the effect of correlations shows up first only in the third-order term.  This effect parallels the exact result in one dimension ($z=2$).

For the long time analysis we follow a similar procedure as for coalescence, this time leading to~\cite{remark}
\begin{equation}
\rho(t)\to1/\big(2\frac{z-2}{z}\,t\big)\;,\qquad\text{as\ }t\to\infty\;.
\end{equation}
Again, in view of the corresponding mean-field equation, $d\rho/dt=-2\rho^2$, the long-time asymptotic result may be understood if the likelihood of pairs of occupied sites is effectively reduced by the factor $(z-2)/z$.

\section*{DISCUSSION}

In conclusion, we have analyzed the kinetics of diffusion-limited coalescence,
\coal, and annihilation, \annil,  in the Bethe lattice, by means of an approximation based on the method of empty intervals.  We find a result that is mean-field in character,
$\rho\sim 1/kt$, where the reaction rate undergoes a slow crossover from $k=1$ (for coalescence) at early times to $k=(z-2)/z$ at late times.  (For annihilation these values are doubled.) The decrease in the reaction rate stems from a buildup of correlations that disfavors particles in close contact.  In the case of annihilation, there is additional behavior: if the initial particle density is larger than $1/2$ then the reaction rate increases above the early mean-field rate, early on, before settling to the lower value of the long-time asymptotic limit.  The buildup of correlations weakens as the coordination number of the graph, $z$, increases, and vanishes altogether in the limit of $z\to\infty$, where the mean-filed limit is regained.  The correlations are strongest 
in the singular limit of $z=2$, corresponding to a linear chain, where the density decay dramatically slows down to a different functional form, $\rho\sim1/\sqrt{t}$~\cite{book}.

Despite the approximation involved, we believe that our results reflect the essential character of the true kinetics.  The long-time behavior can be explained by the following heuristic argument: A particle at shell
$\ell$ from the origin advances to shell $\ell+1$ with probability $(z-1)/z$, and retreats to $\ell-1$ with 
probability $1/z$.  Thus, the particle explores, on average, $(z-1)/z-1/z=(z-2)/z$ new sites per unit time.
Assuming a uniform background of particles, this leads to $d\rho/dt=-[(z-2)/z]k\rho^2$ ($k=1$ for coalescence and $2$ for annihilation), in accord with the approximation results.  (Note, however, that this argument fails to capture the early time behavior predicted by the empty interval approximation.)

Our method relies on translation invariance and does not generalize easily to finite lattices (Cayley trees).  In a finite tree of $L$ shells the particles experience a drift from the center to the rim, which consists of roughly half of the nodes already for $z=3$ (the case relevant to dendrimers).  Thus, the finite-size effect could be substantial and must be reckoned with.  Indeed, in computer simulations  \cite{shapir,kopelman} the density converges quickly to the long-time behavior predicted by our analysis, however, this lasts only until time $t\approx L$ (see, for example, Fig.~5 in \cite{kopelman}),
and thereafter the decay is considerably slower, $\rho\sim 1/t^{0.8}$~\cite{shapir,kopelman}.  Theoretical
analysis of the long-time asymptotics in Cayley trees remains an important open challenge.

\acknowledgments
We thank Paul Krapivsky and Yonathan Shapir for introducing us to the problem, and Drs.~Krapivsky, Shapir and Joseph Skufca for insightful comments and discussions.
Partial support from NSF awards PHY0140094, PHY0555312 (DbA), and DMR0121146 (MLG) is gratefully acknowledged.  

\appendix
\section{Solution of the Rate Equations}
We wish to solve the system of equations
\begin{equation}
\label{En}
\frac{d}{dt}E_n(t)=\frac{n(z-2)+2}{z}\big(E_{n-1}-2E_n+E_{n+1}\big)\,,
\end{equation}
with the boundary conditions
\begin{eqnarray}
\label{E0}
E_0(t)&=&1\,,\\
\label{Einfty}
\lim_{n\to\infty}E_n(t)&=&0\,,
\end{eqnarray}
and the initial condition
\begin{equation}
\label{Einit}
E_n(0)=(1-p)^n.
\end{equation}

First, Laplace-transform in time; 
\[
\phi_n(s)=\int_0^{\infty}e^{-st}E_n(t)\,dt\,,
\]
so that the problem becomes
\begin{equation}
\phi_{n+1}-p_n\phi_n+\phi_{n-1}=c_n\,
\end{equation}
where $p_n=2+\frac{zs}{n(z-2)+2}$, $c_n=-\frac{zE_n(0)}{n(z-2)+2}$,
and $\phi_0=1/s$, $\phi_{\infty}=0$.  Rewrite the underlying homogeneous equation,
$A_{n+1}-p_nA_n+A_{n-1}=0$, as
\begin{equation}
(n+a)A_{n+1}-2(n+b)A_n+(n+a)A_{n-1}=0\,,
\end{equation}
where $a=2/(z-2)$, $b=(2+sz/2)/(z-2)$.
Upon making the substitution
\begin{equation}
A_n=\int_0^1x^{n-1}f(x)\,dx\,,
\end{equation}
the terms linear in $n$ may be reinterpreted as derivatives of $f(x)$ (using integration by parts), and yielding the consistency relation,
\[
\frac{f'(x)}{f(x)}=\frac{2}{(z-2)x}-\frac{zs}{(z-2)(1-x)^2}+\frac{2}{1-x}\,,
\]
leading to 
\begin{equation}
f(x)=\frac{x^{2/(z-2)}}{(1-x)^2}\exp\Big(-\frac{zs}{(z-2)(1-x)}\Big)\,,
\end{equation}
(up to an arbitrary constant factor).  This completes the solution of the homogeneous equation.

To solve the original inhomogeneous equation, make the substitutions
\[
\phi_n=A_nB_n\,,\qquad B_n-B_{n-1}=h_n\,,
\]
yielding
\[
-A_{n-1}h_n+A_{n+1}h_{n+1}=c_n\,.
\]
This has the solution
\begin{eqnarray}
\label{Bn}
B_n&=&\frac{1}{A_0s}+\sum_{k=1}^n h_k\,,\\
h_n&=&\frac{A_0A_1}{A_{n-1}A_n}h_1+\sum_{k=1}^{n-1}\frac{A_k}{A_{n-1}A_n}c_k\,.
\end{eqnarray}
Note that the first term on the RHS
of (\ref{Bn}) guarantees fulfillment of the boundary condition $\phi_0=1/s$.  $h_1$ is determined from the remaining boundary condition, $\phi_{\infty}=0$, and taking into account that $\lim_{n\to\infty}A_n=0$,
\begin{equation}
h_1=-\sum_{k=1}^{\infty}\sum_{l=1}^{k-1}\frac{A_lc_l}{A_{k-1}A_k}/\sum_{k=1}^{\infty}\frac{A_0A_1}{A_{k-1}A_k}\;.
\end{equation}
This concludes the formal solution of the problem.

Manageable expressions may be obtained for the special case of $p=1$ (all lattice sites are occupied, initially), where $c_n=0$, $n\geq1$, leading to $h_1=0$, and $\phi_1=A_1/sA_0$.  Finally, upon evaluating the pertinent integrals, we find for the Laplace transform of the density, ${\hat\rho}(s)=1/s-\phi_1(s)$:
\begin{equation}
{\hat\rho}(s)=\frac{1}{s}\Big(1-\frac{2}{z-2}\frac{U(c,2,cs)}{U(c-1,2,cs)}\Big)\;,
\end{equation}
where $c=z/(z-2)$ and $U(\cdot)$ is Kummer's confluent hypergeometric function~\cite{AS}.  Inversion of this expression in the  limits of short and long times agrees with the results derived in the paper by independent methods.


\begin{thebibliography}{99}
\bibitem{remark}  The scaling function, analogous to $\Phi(\xi)$ of Eq.~(\ref{Phi}), is now
$1/2+(1/2)\exp[-z\xi/(z-2)]$.
\end{thebibliography}
\end{document}